\documentclass[11pt,a4paper]{amsart}
\usepackage[margin=2.5cm]{geometry}
\usepackage{hyperref}
\usepackage{amsfonts,amssymb,amsmath,graphicx}

\newtheorem{theorem}{Theorem}[section]
\newtheorem{example}[theorem]{Example}
\newtheorem{remark}[theorem]{Remark}

\usepackage{tikz,xcolor}

\definecolor{lime}{HTML}{A6CE39}
\DeclareRobustCommand{\orcidicon}
{
	\begin{tikzpicture}
	\draw[lime, fill=lime] (0,0) 
	circle [radius=0.17] 
	node[white] {{\fontfamily{qag}\selectfont \tiny ID}};
	\draw[white, fill=white] (-0.0625,0.095) 
	circle [radius=0.007];
	\end{tikzpicture}
	\hspace{-2mm}
}
\foreach \x in {A, ..., Z}
{\expandafter\xdef\csname orcid\x\endcsname
	{\noexpand\href{https://orcid.org/\csname orcidauthor\x\endcsname}
		{\noexpand\orcidicon}
	}
}

\title{A new operational matrix technique to solve linear boundary value problems}

\author{\textbf{Udaya Pratap Singh\hspace{-0.5em} \orcidA{}}\\
	Department of Applied Sciences \\Rajkiya Engineering College, Sonbhadra, Uttar Pradesh, India\\ 
	\textit{email}: \textit{upsingh1980@gmail.com}\\
	Orcid \hspace{-0.6em}\orcidA{}: \url{https://orcid.org/0000-0002-4538-9377}}
\begin{document}
\maketitle
\begin{abstract} A new technique is presented to solve a class of linear boundary value problems (BVP). Technique is primarily based on an operational matrix developed from a set of modified Bernoulli polynomials. The new set of polynomials is an orthonormal set obtained with Gram-Schmidt orthogonalization applied to classical Bernoulli polynomials. The presented method changes a given linear BVP into a system of algebraic equations which is solved to find an approximate solution of BVP in form of a polynomial of required degree. The technique is applied to four problems and obtained approximate solutions are graphically compared to available exact and other numerical solutions. The method is simpler than many existing methods and provides a high degree of accuracy.
	

\end{abstract}

\section{Introduction}
\label{section 1}
Boundary value problems (BVP) have a lot of applications in areas of science, engineering and technology. For illustration, rheological models, bio-fluid models, industrial engineering, hydrodynamics, lubrication problems, economics, ecology models, biological models, heat and mass transfer and many more are the examples where the BVPs naturally arise to play a significant role. It is often hard to find an analytic solution to these BVPs. In such situations, an approximate or numerical solution becomes an essential tool to deal with the problems. Investigations of numerical schemes to solve BVPs have been of concern from long past \cite{Keller1968,Greenspan1988,Miller1996}, however, in $20^{th}$ century, the advent of modern computers and software attracted much attention of researchers towards high precision computations to numerical approximation problems \cite{Yousefi2006,Xu2007,Bhrawy2012}, which has been of major concern in present times due to increasing demand of high precision numerical solutions in different fields \cite{Iqbal2018,Shiralashetti2019, Samadyar2019}. Some notable works on numerical or approximate  solutions of BVPs also include \cite{Cheng2005, Valarmathi2010,Lang2012,Ramos2020}. Many authors used different polynomials such as Chebyshev polynomials \cite{Maleknejad2007}, Legendre polynomials \cite{Nemati2015}, Laguerre polynomials and Wavelet Galerkin method \cite{Rahman2012}, Legendre wavelets \cite{Yousefi2006} to present various numerical schemes. Bernoulli polynomials and its properties have also been taken into account by many researchers \cite{Cheon2003,Natalini2003,Kurt2011,Mohsenyzadeh2016}. Recently, Singh et al. [20] used Bernoulli polynomials to solve Abel-Volterra type integral equations. However, numerical schemes always provide a numerical solution, it may not qualify for further analytical applications in various situations. Therefore, the need of a precise and simple approximate solution is always motivated. 

It is, therefore, proposed to solve linear boundary value problems of ordinary differential equations using a class of modified Bernoulli polynomials and an operational matrix thereof to find an approximate solution in the form of a polynomial.
\section{Modified Bernoulli Polynomials}
\label{section 2: Bernoulli Polynomials }
Classical Bernoulli polynomials are given as \cite{Costabile2006}:
\begin{equation}\label{eq.1 : Basic Bernoulli Polynomials}
{B_n}(\zeta ) = \sum\limits_{j = 0}^n {\,\left( {\begin{array}{*{20}{c}}
		n \\ j 	\end{array}} \right)\,\,{\acute{b}_j}\,{\zeta ^{n - j}},\hspace{1em} n = 0,1,2,...} \hspace{1em} 0 \le \zeta  \le 1
\end{equation}
where, $\acute{b}_j$ are the Bernoulli numbers, which can be easily calculated with Kronecker’s formula \cite{Todorov1984}:
\begin{equation} \label{Kronecker’s formula}
{B_n}(0) =  - \sum\limits_{j = 1}^{n + 1} {\frac{{{{( - 1)}^j}}}{j}\left( {\begin{array}{*{20}{c}}
		{n + 1}\\j \end{array}} \right)} \sum\limits_{k = 1}^j {{k^n}} \,\,;\,\,\,n \ge 0
\end{equation} 

For illustration, expanded expression for first four Bernoulli polynomials are : ${B_0}(\zeta) = 1, B_1(\zeta)=\zeta-\frac{1}{2}, B_2(\zeta)=\zeta^2-\zeta+\frac{1}{6}, B_3(\zeta)=\zeta^3-\frac{3}{2}\zeta^2+\frac{1}{2}\zeta, 	B_4(\zeta)=\zeta^4-2\zeta^3+\zeta^2-\frac{1}{30}$.

Many interesting properties of Bernoulli polynomials have been studied by different researchers from time to time \cite{Costabile2001, Kreyszig1978}. Two of its properties that are of interest in the present work are that these polynomials form a complete basis over $[0,1]$ \cite{Kreyszig1978}, and their integral over $[0,1]$ is uniformly zero \cite{Costabile2001},
	\begin{equation}\label{Bernouli Integral 0}
	\int_0^1 {{B_n}(z)dz = 0, \: n \ge 1}.
	\end{equation}
	
Some other properties such as:	
	\begin{equation}
	\begin{split}
	B'_n(\zeta ) = n{B_{n - 1}}(\zeta ),\: n \ge 1,\\
	{B_n}(\zeta  + 1) - {B_n}(\zeta ) = n{\zeta ^{n - 1}},\: n \ge 1
	\end{split}
	\end{equation}
and many more including their generalization and advanced applications can be found in notable literature  \cite{Natalini2003,Costabile2006,Kurt2011,Lu2011}.

\subsection{Gram-Schmidt orthogonalization}
\label{section 3: The Orthonormal Polynomials}
Property (\ref{Bernouli Integral 0}) shows that the polynomials ${B_n}(\zeta)\,(n \ge 1)$ (\ref{eq.1 : Basic Bernoulli Polynomials})  are orthogonal to $B_o(\zeta)$ with respect to standard inner product on  ${L^2} \in [0,1]$ defined as:
\begin{equation}\label{inner product}
<f_1,f_2> = \int_{0}^{1} f_1(x) \bar{f_2(x)} dx \,; \:\: f_1,f_2 \in L^2[0,1]
\end{equation}

With inner product (\ref{inner product}), an orthonormal set of $n+1$ polynomials is derived for any $B_n$ with Gram-Schmidt orthogonalization. For illustration, $n=5$ gives following set of modified orthonormal polynomials:
\begin{equation} \label{Orthornormal polynomials}
\begin{split}
{\phi _{0\,}}(\zeta ) &= 1\\
{\phi _1}(\zeta ) &= \sqrt 3 ( - 1 + 2\zeta )\\
\phi_2\left(\zeta\right)&=\sqrt5\left(1-6\zeta+6\zeta^2\right)\\
\phi_3(\zeta)&=\sqrt7(-1+12\zeta-30\zeta^2+20\zeta^3)\\
\phi_4(\zeta)&=3(1-20\zeta+90\zeta^2-140\zeta^3+70\zeta^4)\\
\phi_5(\zeta)&=\sqrt{11}(-1+30\zeta-210\zeta^2+560\zeta^3-630\zeta^4+252{\zeta^5})
\end{split}
\end{equation}

\subsection{Operational matrix}
\label{section 5 : Construction of operational matrix}
On integration over the interval $[0,1]$, the orthonormal polynomials for (\ref{Orthornormal polynomials}) shows following relation:
\begin{equation}\label{eq.7 : Ortho Pol. of deg. 0}
\int_0^\zeta  {{\phi _o}(\eta )d\eta }  = \frac{1}{2}{\phi _o}(\zeta ) + \frac{1}{{2\sqrt 3 }}{\phi _1}(\zeta )
\end{equation}

\begin{equation}\label{eq.8 : Ortho Pols.}
\begin{array}{l}
\int\limits_0^\zeta  {{\phi _i}(x)dx = } \,\,\,\,\,\frac{1}{{2\sqrt {(2i - 1)(2i + 1)} }}{\phi _{i - 1}}(\zeta )\\
\hspace{2cm} + \frac{1}{{2\sqrt {(2i + 1)(2i + 3)} }}{\phi _{i + 1}}(\zeta ),\,\,\,\,(\,for{\rm{ }}\,i = 1\,,2,...\,,n)
\end{array}
\end{equation}
which can be represented in following closed form:
\begin{equation}\label{Integral of Ortho pol}
\int\limits_0^\zeta  {\phi(\eta )d\eta  = \,\,} \Theta \,\phi(\zeta )
\end{equation}

where $\zeta\in[0,1]$ and $\Theta$ is operational matrix of order $(n+1)$ given as :
\begin{equation} \label{Operational matrix}
\Theta \, = \frac{1}{2}\left[ {\begin{array}{*{20}{c}}
	1&{\frac{1}{{\sqrt {1.3} }}}&0& \cdots &0\\
	{ - \frac{1}{{\sqrt {1.3} }}}&0&{\frac{1}{{\sqrt {3.5} }}}& \cdots &0\\
	0&{ - \frac{1}{{\sqrt {3.5} }}}&0& \ddots & \vdots \\
	\vdots & \vdots & \ddots & \ddots &{\frac{1}{{\sqrt {\left( {2n - 1} \right)\left( {2n + 1} \right)} }}}\\
	0&0& \cdots &{ - \frac{1}{{\sqrt {\left( {2n - 1} \right)\left( {2n + 1} \right)} }}}&0
	\end{array}} \right]
\end{equation}

\section{Solution of Boundary Value Problems}
\subsection{Approximation of Functions}
\label{section 4 : Approximation of Functions}

\begin{theorem} \label{theorem 1}
	 Let $H=L^2[0,1]$  be a Hilbert space and $Y=span\left\{y_0,y_1,y_2,...,y_n\right\}$  be a subspace of $H$ such that $dim{(}Y)<\infty$ , every $f\in H$  has a unique best approximation out of $Y$ \cite{Kreyszig1978}, that is, $\forall y(t)\in Y,\, \exists \, \hat{f}(t)\in Y$ s.t. $\parallel f(t)-\hat{f}(t)\parallel_2\le\parallel f(t)-y(t)\parallel_2$. This implies that,  $\forall \, y(t)\in Y, <f(t)-f(t), y(t)>= 0$, where $<,>$  is standard inner product on $L^2\in[0,1]$ (\textit{c.f. Theorems 6.1-1 and 6.2-5, Chapter 6} \cite{Kreyszig1978}).
\end{theorem}

\begin{remark}
	Let $Y=span\left\{\phi_0,\phi_1,\phi_2,...,\phi_n\right\},$ where $\phi_k\in L^2[0,1]$ are orthonormal Bernoulli polynomials. Then, from Theorem \ref{theorem 1}, for any function $ f\in L^2[0,1],$
	\begin{equation} \label{eq.5 : approx theorem}
	f\approx\hat{f}=\sum_{k=0}^{n}{c_k\phi_k}, 
	\end{equation}
	where $c_k=\left\langle f,\phi_k\right\rangle,$ and $<,>$ is the standard inner product on $L^2\in[0,1]$ as defined by equation (\ref{inner product}).
\end{remark}

For numerical approximation, series (\ref{eq.5 : approx theorem}) can be written as:
\begin{equation} \label{eq.6 : approximation series}
f(\zeta)\simeq\sum_{k=0}^{n}{c_k\phi_k=C^T\phi(\zeta)} 
\end{equation}
where $C=\left(c_0,c_1,c_2,...,c_n\right), \phi(\zeta)=\left(\phi_0,\phi_1,\phi_2,...,\phi_n\right)$ are column vectors, and number of polynomials $n$ can be chosen to meet required accuracy.

\subsection{Scheme of Approximation}
\label{section 6: Solution of Initial Value Problems}
In order to present the basic ingredients of the method in simpler way,  general form of second order linear ordinary differential equation with constant coefficients will be considered first; and application of the method to higher order BVPs of similar kind will be discussed subsequently.

Let us consider the linear ordinary differential equation with constant coefficients:
\begin{equation}\label{ODE 2nd order}
\frac{{{d^2}y}}{{d{\zeta ^2}}} + a_1\,\frac{{dy}}{{d\zeta }} + a_0\,y = r(\zeta )
\end{equation}

Without loss of generality, we assume that the ODE (\ref{ODE 2nd order}) satisfy the boundary conditions (BCs):
\begin{equation} \label{BCs 0,1}
\:\:y(0) = \alpha, \: y(1) = \beta
\end{equation}
for if the boundary conditions be $y(\zeta_0)=\alpha, y(\zeta_1)=\beta$, BCs (\ref{BCs 0,1}) can be attained with the transformation $\zeta\rightarrow \frac{\zeta-\zeta_0}{\zeta_1-\zeta_0}$. It is further assumed that $y$ and $r$ are continuous functions of $\zeta\in[0,1]$ and BVP (\ref{ODE 2nd order}-\ref{BCs 0,1}) admits a unique solution on $[0,1]$.

Let $C=\left(c_o,c_1,c_2,...,c_n\right)$ be a column vector of $n+1$ unknown quantities such that
\begin{equation}\label{y''=CT phi}
\frac{{{d^2}y}}{{d{\zeta ^2}}} = C^T\,\phi(\zeta )
\end{equation}

Equations (\ref{y''=CT phi}) and (\ref{Integral of Ortho pol}-\ref{Operational matrix}) give:
\begin{equation}\label{dy/dx and y}
\begin{split}
\frac{dy}{d\zeta} &= \beta-\alpha-C^T \Theta^2 \phi(1) + C^T \Theta \phi(\zeta)\\
y(\zeta) &= \alpha + (\beta-\alpha-C^T \Theta^2 \phi(1)) \zeta + C^T \Theta^2 \phi(\zeta)
\end{split}
\end{equation}
where, $\phi(1)=\left(1,\sqrt{3},\sqrt{5},\dots, \sqrt{2n+1}\right)^T$.

Substituting equations (\ref{y''=CT phi}-\ref{dy/dx and y}) into ODE (\ref{ODE 2nd order}), we get:
\begin{equation}\label{LHS Simplified of ODE}
C^T\left(I + a_1\Theta +a_0 \Theta^2\right){\phi}(\zeta ) - C^T \left(a_0\zeta+a_1\right)\Theta^2\phi(1)= r(\zeta )
\end{equation}
where, $I$ is identity matrix of order $n+1$. Again, writing  
\begin{equation}
\left(a_0\zeta+a_1\right)\Theta^2\phi(1) = \mathbb{L} \phi(\zeta)
\end{equation}
and 
\begin{equation} \label{R^T phi}
r(\zeta) - (\beta-\alpha)(a_0\zeta+a_1) - a_0\alpha = R^T\phi(\zeta),
\end{equation}
equation (\ref{LHS Simplified of ODE}) is simplified to following form:
\begin{equation}\label{ODE fully simplified}
C^T\left(I + a_1\Theta +a_0 \Theta^2 + \mathbb{L}\right){\phi}(\zeta ) = R^T \phi(\zeta)
\end{equation}
where, $R=\left(r_o,r_1,...,r_n \right)$ is a real column vector and $\mathbb{L}$ (for this case) is calculated as,
\begin{equation}
	\mathbb{L} = \left( {\begin{array}{*{20}{c}}
		{\frac{1}{4}({a_0} + 2{a_1})}&{\frac{1}{{4\sqrt 3 }}{a_o}}&0& \cdots &0\\
		{ - \frac{{\sqrt 3 }}{{12}}({a_0} + 2{{\rm{a}}_{\rm{1}}})}&{ - \frac{1}{{12}}{a_0}}&0& \cdots &0\\
		0&0&0& \cdots &0\\
		\vdots & \vdots & \ddots & \ddots & \vdots \\
		0&0&0& \cdots &0
		\end{array}} \right)_{(n+1)\times(n+1)}
\end{equation}

From equation (\ref{ODE fully simplified}) and (\ref{y''=CT phi}), the unknown coefficient $C^T$ and approximate solution to BVP (\ref{ODE 2nd order}-\ref{BCs 0,1}) are obtained as:
\begin{equation}\label{C^T for ODE}
C^T = R^T \left(I + a_1\Theta +a_0 \Theta^2 + \mathbb{L}\right)^{-1}
\end{equation}
\begin{equation} \label{Soln. for ODE}
y(\zeta ) = \alpha +\left(\beta-\alpha-C^T \Theta^2 \phi(1)\right)\zeta + C^T \,\Theta^2\,{\phi}(\zeta ).
\end{equation}

\begin{remark}
	For the shake of completeness, let the BVP under consideration be of order $n$. Following obvious management will be required to the intermediate steps:
\end{remark}
	\begin{itemize}
	\item $n^{th}$ derivative will be set to $C^T\phi$ (say, $\frac{d^ny}{d\zeta^n} = C^T\phi $).
	\item left hand side of equation (\ref{R^T phi}) will be adjusted with the polynomial generated due to $n$ number of $BC_S$.
	\item equation (\ref{C^T for ODE}) will take the form $C^T =  R^T \left(I + a_{n-1}\Theta +a_{n-2} \Theta^2 +\dots +a_0 \Theta^n + \mathbb{L}\right)^{-1}$ with appropriately calculated $\mathbb{L}$.
	\end{itemize}

\section{Examples}
\label{section 7 : Examples}
In this section, four examples have been considered to demonstrate the efficacy of the method. The first example is taken to demonstrate the scheme of approximation, second and fourth examples are taken from published investigations, and the third example is selected for the reason that it has no easy analytic solution. 

\begin{example} \label{Example 1}
	As a first example, let us consider the following simple boundary value problem:  
	
	\begin{equation} \label{Ex.1}
	\begin{split}
	\frac{d^2 y}{\text{dx}^2}-5\frac{\text{dy}}{\text{dx}}+6 y = e^{-x};\:\:y(0)=0,\:\: y(1)=5
	\end{split}
	\end{equation}
	which has exact solution $y(x) =  \frac{1}{12}\left(e^{-x}-\frac{\left(e^4+60 e-1\right)}{e^3 (e-1)}e^{2 x}+\frac{\left(e^3+60 e-1\right)}{e^3 (e-1)} e^{3 x}\right)$.
\end{example}

Comparing equation (\ref{Ex.1}) with equation (\ref{ODE 2nd order}) for $n=7$ ans $10$, equations (\ref{C^T for ODE} - \ref{Soln. for ODE}) yield
\begin{equation}\label{C^T for Ex 1}
\begin{split}
C^T_{n=7} = \left(18.5536,\, 15.4731,\, 6.0611,\, 1.5558,\, 0.296729,\, 0.044957,\, 0.00571811,\, 0.000619857\right)\\
C^T_{n=10} = \left(18.5536,\, 15.4731,\, 6.06111,\, 1.5558,\, 0.296729,\, 0.0449654,\, 0.00565359,\, 0.000607531,\, \right.\\
\left. 0.0000570205,\, 4.75158\times 10^{-6},\, 3.57786\times 10^{-7}\right)
\end{split}
\end{equation}
\begin{equation} \label{y(x) for Ex 1}
\begin{split}
y(x)_{n=7} \approx  0.190686 x + 0.953731 x^2 + 1.44662 x^3 + 0.441934 x^4 + 2.15981 x^5 - 1.08742 x^6 + 0.894632 x^7\\
y(x)_{n=10} \approx  -3.489\times10^{-7} + 0.1899319 x + 0.973711859 x^2 + 1.27968935 x^3 + 
1.068523372 x^4 \\
+ 1.00141756 x^5 - 0.1468994867 x^6 + 0.912766741 x^7 - 0.5236070455 x^8 \\
+ 0.2778958181 x^9 - 0.0334297252 x^{10}
\end{split}
\end{equation}

\begin{figure}[h!]
\centering
\includegraphics[width=0.5\linewidth]{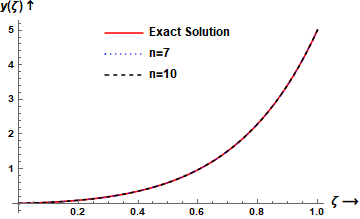}\vspace{1mm}\\
(a) \vspace{2mm}\\
\includegraphics[width=0.48\linewidth]{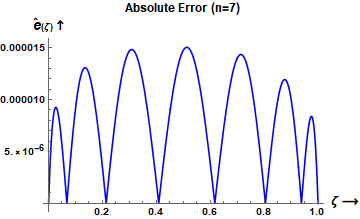} 
\includegraphics[width=0.48\linewidth]{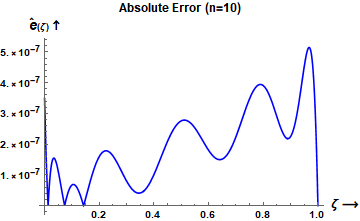}\\ \vspace{2mm}
(b)
\caption{(a) Comparison of exact and present solution for example \ref{Example 1} for $n=7,10$. (b) Absolute error between exact and approximate solutions of  \ref{Example 1} for $n=7$ and $10$.}
\label{fig1}
\end{figure}

A comparison of approximations (\ref{y(x) for Ex 1}) with exact solution of IVP (\ref{Ex.1}) is shown in figure \ref{fig1}. Maximum magnitude of the error between exact and present solutions is of order $10^{-5}$ and $10^{-7}$ for $n=7$ and $n=10$, respectively. It is notable that the error for $n=10$ is equivalent to error of concatenated series of exact solution at degree $15$.

\begin{example} \label{Example 2}
	Let us consider the following boundary value problem of order $9$ \cite{WAZWAZ2000}:  
	\begin{equation} \label{Ex.2}
	\begin{split}
	\frac{d^9 y}{\text{dx}^9} - y = -9e^x \\
	y^{(k)}(0)=0,\:\: k = 0,1,2,3,4\\
	y^{(k)}(1)= - k\, e,\:\: k = 0,1,2,3
	\end{split}
	\end{equation}
	where, $y^{(k)}(x) = \frac{d^k y}{\text{dx}^k}$. This BVP admits the exact solution $y(x) = (1 - x) e^x$.
\end{example}

Wazwaz \cite{WAZWAZ2000} presented an approximate solution of BVP (\ref{Ex.2}) as polynomial of degree $12$ and got an error of order $10^{-9}$. In order to compare our solution to that by Wazwazz \cite{WAZWAZ2000}, we will present an approximation of degree 7 and 12. 
\begin{figure}[h!]
	\centering
	\includegraphics[width=0.48\linewidth]{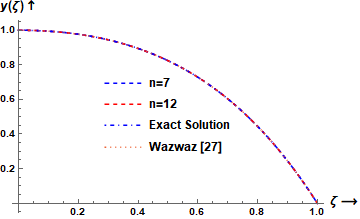}\vspace{1mm}
	\includegraphics[width=0.48\linewidth]{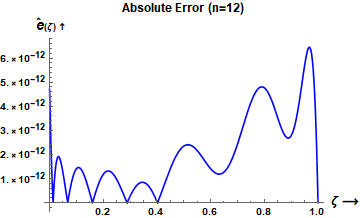} \\
	(a) \hspace{0.5\linewidth} (b)\\ \vspace{5mm}
	\includegraphics[width=0.48\linewidth]{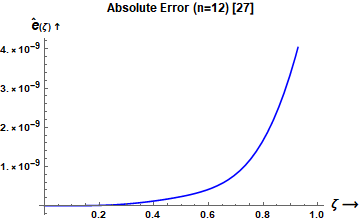}	\includegraphics[width=0.48\linewidth]{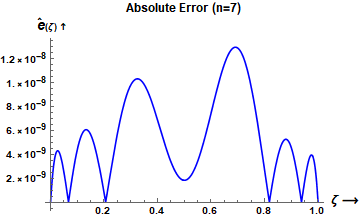}\\ 
	(c) \hspace{0.5\linewidth}(d)
	\caption{(a) Comparison of exact and present solution to example \ref{Example 2} for $n=7,12$. (b) Absolute error between exact and approximate solutions for $n=7$ (present approximation). (c) Absolute error between exact and approximate solutions \cite{WAZWAZ2000}. (d) Absolute error between exact and approximate for $n=12$ (present approximation).}
	\label{fig2}
\end{figure}

Proceeding with the BVP (\ref{Ex.2}) as discussed in section \ref{section 6: Solution of Initial Value Problems} for  $n (n=7, 12)$, we get:
\begin{equation}
\begin{split}
C^T_{n=7} = \left( -16.1828, -4.12338, -0.466916, -0.0340935, -0.00181803,  -0.0000761389, \right.\\
			\left. -2.77032\times 10^{-6}, -1.00487\times  10^{-7} \right)\\
C^T_{n=12} = \left( -14.7463, -4.65975, -0.65899, -0.0606994, -0.00414329, -0.000224385,\right.\\
-0.0000100613, -3.84638\times  10^{-7}, -1.28069\times  10^{-8}, -3.77223 \times 10^{-10},\\
\left. -3.55521\times  10^{-13}, -2.28925\times  10^{-13}, -4.33444\times  10^{-15}  \right)
\end{split}
\end{equation}
\begin{equation} \label{y(x) for Ex 2}
\begin{split}
y(x)_{n=7} \approx  1 - 3.72702\times  10^{-7} x - 0.49999 x^2 - 0.333419 x^3 - 0.124664 x^4 - 
0.0340174 x^5\\ - 0.00621597 x^6 - 0.00154015 x^7 - 0.000152922 x^8\\
y(x)_{n=12} \approx  -3.489\times10^{-7} + 0.1899319 x + 0.973711859 x^2 + 1.27968935 x^3 + 
1.068523372 x^4 \\
+ 1.00141756 x^5 - 0.1468994867 x^6 + 0.912766741 x^7 - 0.5236070455 x^8 + 0.2778958181 x^9 \\- 0.0334297252 x^{10}-1.9738441\times 10^{-7} x^11 - 3.9815891\times 10^{-8} x^12
\end{split}
\end{equation}

Approximations (\ref{y(x) for Ex 2}) are compared with exact and approximate solution \cite{WAZWAZ2000} of BVP (\ref{Ex.2}) in figure \ref{fig1}. Maximum magnitude of the error between exact and present solutions is of order $10^{-8}, 10^{-12}$ for $n=7$ and $n=12$, respectively. It is notable that the error of approximation of $12^{th}$ degree polynomial by Wazwaz \cite{WAZWAZ2000} is of order $10^{-9}$, which is closer to that for $n=7$ of present solution, whilst our solution for $n=12$ is much more accurate than Wazwaz \cite{WAZWAZ2000}.

\begin{example} \label{Example 4}
	Consider the ODE  
	\begin{equation} \label{Ex.4}
	\frac{d^2y}{dx^2}-5 \frac{dy}{dx}+2 y = \tan(x) ; \hspace{1em} y(0)=\left( \frac{dy}{dx}\right) _{x=0} =0
	\end{equation}
	which is linear in nature but its not easy to solve manually in terms of simply known mathematical functions. We will compare the present solution of this problem with numerical solutions generated by Mathematica.
\end{example}

Proceeding as in previous examples for $n=9,11$, we get
\begin{equation}\label{CT for Ex.2}
{C^T} = \left( 5.1220,\, 5.5181,\, 2.9304,\, 1.0668, \, 0.2958,\, 0.0663, \, 0.0125,\,  0.0020, \, 0.0003 \right)
\end{equation}
\begin{equation}\label{y(x) for Ex 4}
\begin{split}
y(x) \approx 0.0001 x - 0.0025 x^2 + 0.1942 x^3 + 0.04799 x^4 + 0.7521 x^5 - 0.9599 x^6 \\+ 1.5043 x^7 - 0.9351 x^8 + 0.3669 x^9
\end{split}
\end{equation}

The approximate solution (\ref{y(x) for Ex 4}) is compared with exact solution of IVP (\ref{Ex.4}) and observed absolute error of orders $10^{-4},\, 10^{-5}$ for $n=9,\,11$, respectively (Figure \ref{fig2}).  

\begin{figure}[h!]
	\centering
	\includegraphics[width=0.48\linewidth]{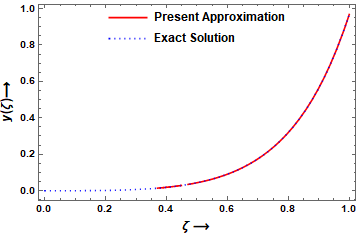} 
	\includegraphics[width=0.48\linewidth]{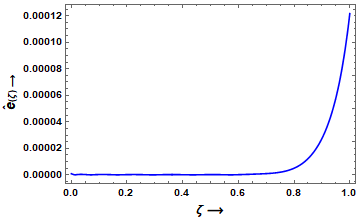}\\
	(a) \hspace{0.5\linewidth} (b)
	\caption{(a) Comparison of present approximation and $Mathematica$ generated numerical solutions to example \ref{Example 2} for $n=9,\,11$. (b) Absolute error between present approximation and $Mathematica$ generated numerical solutions for $n=9,\,11$.}
	\label{fig4}
\end{figure}

\begin{example} \label{Example 3}
	As a last example, let us take the following BVP \cite{Barari2008} of order four:  
	\begin{equation} \label{Ex.3}
	\begin{split}
	\frac{d^4y}{dx^4}-\frac{d^2y}{dx^2} - y &= (x-3) e^x \\
	y(0)= (1), \: y(1) = 0, \: &\left( \frac{dy}{dx}\right) _{x=0} =0, \: \left( \frac{dy}{dx}\right) _{x=1} = -e
	\end{split}
	\end{equation}
	which is admits the exact solution $y(x)=(1-x)e^x$.
\end{example}

 Barari et al. \cite{Barari2008} presented an approximate solution of degree $11$ with variational iteration method (VIM) and got an error of order $10^{-5}$. We have presented solutions for $n=7,10$ and obtained errors of order $10^{-5}$ and $10^{-8}$, respectively.
 
Proceeding as in previous examples for $n=7,10$, we obtained
\begin{equation}\label{CT for Ex.3}
\begin{split}
{C^T}_{(n=7)} = \left(\right. 0.7182771032, -0.26820503188, -0.0960498415, -0.0133080241, -0.00116353896,\\ -0.0000739287, -3.481050564\times 10^{-6}, -9.8519112388\times 10^{-8} \left. \right)\\
{C^T}_{(n=10)} =\left(\right. 0.718281826, -0.26820025078, -0.0960487105, -0.0133096959, -0.00116552426,\\ -0.00007500648, -3.8229361187\times 10^{-6},-1.6126955267\times 10^{-7},\\ -5.755809746\times 10^{-9},-1.68985029\times 10^{-10}, 3.2515786\times 10^{-12} \left. \right)
\end{split}
\end{equation}
\begin{equation}\label{y(x) for Ex 3}
\begin{split}
y(x)_{(n=7)} \approx 0.9999999999  - 0.5 x^2 - 0.333333 x^3 - 0.125 x^4 - 0.0333433 x^5 \\- 0.00701389 x^6 - 0.00130952 x^7\\
y(x)_{(n=10)} \approx 0.9999999999 - 0.5 x^2 - 0.33333333 x^3 - 0.125 x^4 - 
0.0333333 x^5 \\- 0.00694444 x^6 - 0.00119048 x^7 - 0.000173611 x^8 \\- 
0.000022048 x^9 - 2.75298083 \times 10^{-6} x^{10}
\end{split}
\end{equation}

\begin{figure}[h!]
	\centering
	\includegraphics[width=0.6\linewidth]{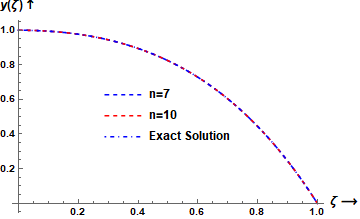}\vspace{1mm}\\
	(a) \\ \vspace{5mm}
	\includegraphics[width=0.48\linewidth]{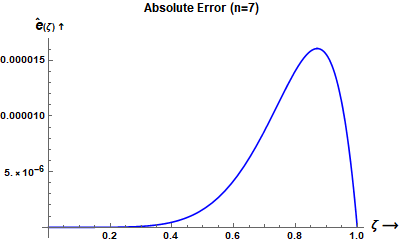}	\includegraphics[width=0.48\linewidth]{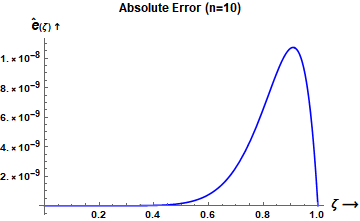}\\ 
	(b) \hspace{0.5\linewidth}(c)
	\caption{(a) Comparison of exact and present solution for example \ref{Example 3} for $n=7,10$. (b) Absolute error for $n=7$ . (c)  Absolute error for $n=10$ .}
	\label{fig3}
\end{figure}

Figure \ref{fig3} shows comparison of present approximation with exact solution to example \ref{Example 3} for $n=7, 9$. It is easy to observe that the present method for $n=7$ yields similar error as in \cite{Barari2008}, but our solution for $n=10$ yields far better approximation than that obtained in \cite{Barari2008}. If value of $n$ is taken higher, more accurate solution will be obtained.

\section{Conclusion}
A new scheme was presented and demonstrated to approximate the solution of linear boundary value problems with constant coefficients. Gram-Schmidt orthogonalization and standard inner product of $L^2[0,1]$ applied to a set of first $n$ Bernoulli polynomials produced a new class of $n$ orthonormal polynomials showing a special tri-diagonal operational matrix, which were utilized as a tool to transform a BVP into a system of algebraic equations with unknown coefficients. These unknown coefficients are evaluated with the scheme discussed in present method and, thereby, a polynomial approximation to the solution of the BVP is  obtained. The method was explored with three examples. The main benefits of this method can be concluded as follows:

\begin{itemize}
	\item approximate solution comes out to be a polynomial of degree $n$, which enables the further application of solution.  
	\item approximation contain small errors, which can be minimized by considering higher degree of Bernoulli polynomials.
	\item method is fast in comparison to many available numerical and approximation methods.  
\end{itemize}

\bibliographystyle{ieeetr}
\bibliography{manuscript}

\end{document}